\documentclass[12pt,a4paper]{article}
\usepackage{psfig}  
\renewcommand{\thefootnote}{\fnsymbol{footnote}}
\newcounter{line}

\def\ie{\hbox{\it i.e.}{}}
\def\eg{\hbox{\it e.g.}{}}
\def\etc{\hbox{\it etc}{}}
\def\nn{\hspace{2mm}}
\def\sss{\scriptscriptstyle}
\newcommand{\MeV}{\mbox{\rm MeV}}
\newcommand{\GeV}{\mbox{\rm GeV}}
\newcommand{\eV}{\mbox{\rm eV}}
\def\sleq{\raisebox{-.6ex}{${\textstyle\stackrel{<}{\sim}}$}}
\def\sgeq{\raisebox{-.6ex}{${\textstyle\stackrel{>}{\sim}}$}}
\def\Tr{{\rm Tr}{}}
\def\Tilde#1{\widetilde{#1}}

\def\sVEV#1{\left\langle #1\right\rangle}

\def\abs#1{\left| #1\right|} 
\def\cL{{\cal L}}
\def\AGUT{{}\;\;\raisebox{.9ex}{$\times$}\raisebox{-.5ex}%
{$\!\!\!\!\!\!\!\!\sss i=1,2,3$} \,(SMG_i \times U(1)_{\sss B-L,i})}%

\begin{document}
\begin{titlepage}
\begin{flushleft}
\vspace{-1.5cm}
\vbox{
    \halign{#\hfil        \cr
           DESY 02-084    \cr
           NBI-HE-02-08    \cr
           hep-ph/0207023  \cr
           July 2002 \cr
           } 
      }  
\end{flushleft}
\vspace*{-1cm}
\begin{center}
{\Large {\bf Non-thermal Leptogenesis from the Heavier 
Majorana Neutrinos}}

\vspace*{7mm}
{\ T. Asaka}$^{\it a,}$\footnote[1]{E-mail: asaka@mail.desy.de}, 
{\ H. B. Nielsen}$^{\it a,b,}$\footnote[2]{E-mail: hbech@mail.desy.de} 
and {\ Y. Takanishi}$^{\it a,b,}$\footnote[3]{E-mail: yasutaka@mail.desy.de}
            
\vspace*{.2cm}
{\it $^a$ Deutsches Elektronen-Synchrotron DESY, \\
Notkestra{\ss}e 85,\\
D-22603 Hamburg, \\
Germany}\\
\vskip .3cm
{\it $^b$ The Niels Bohr Institute,\\
Blegdamsvej 17, \\ 
DK-2100 Copenhagen {\O}, \\
Denmark}\\
\vspace*{.3cm}
\end{center}
\begin{abstract}
We investigate a scheme for making leptogenesis by means of the $CP$
violating decays of the seesaw Majorana neutrinos proposed by
Fukugita and Yanagida. However, in order to avoid the wash-out of
the produced lepton number we propose the production of the Majorana
neutrinos to occur {\em non-thermally} and sufficiently {\em late}. After this
time, in consequence, the $B-L$ (baryon minus lepton) quantum 
number becomes a good ``accidental symmetry'' protecting the asymmetry 
produced.
This non-thermal leptogenesis at late time is realized by a boson
decaying into the Majorana neutrinos with a long lifetime.
Suggestively this boson could correspond to a scalar field which causes the
cosmic inflation, the inflaton, and thus its decay
means really the reheating of the Universe.
We find that this mechanism works well
even if the lightest Majorana neutrinos are not produced
sufficiently or not present, and the decays of the {\em heavier} seesaw
Majorana neutrinos can be responsible to the baryon asymmetry in the
present Universe, as we illustrate by the example of the family 
replicated gauge group model.

\vskip 4.5mm \noindent\
\vskip -3mm \noindent\
\end{abstract}

\end{titlepage}
\newpage
\renewcommand{\thefootnote}{\arabic{footnote}}
\setcounter{footnote}{0}
\setcounter{page}{1}
\section{Introduction}

Matter-antimatter asymmetry in the present Universe is one of the
biggest puzzles in particle physics as well as in cosmology. This baryon
asymmetry is usually expressed by the ratio of baryon (minus anti-baryon)
number density $n_B$ to the entropy density $s_0$ in the present Universe
as~\cite{Groom:in}
\begin{eqnarray}
  \label{BA}
  \frac{n_B}{s_0} = ( 3.7 - 8.8 ) \times 10^{-11}\nn.
\end{eqnarray}
It is only one, but mysterious, number in nature which we would like to
understand. If our Universe experienced the inflationary stage in the
beginnings of the history, the primordial baryon asymmetry would be
diluted away and be essentially zero. The observed asymmetry in
Eq.~(\ref{BA}) should, therefore, be generated after the inflation. Such a
generation mechanism is called baryogenesis and various 
scenarios have been proposed so far.

Evidence of the neutrino oscillations gives an important clue for
baryogenesis, since Fukugita and Yanagida~\cite{Fukugita:1986hr}
proposed that lepton-number violation in nature might account for the
present baryon asymmetry. The experimental data suggest tiny but
non-zero masses for the neutrinos. Introducing the right-handed Majorana
neutrinos having heavy masses is the natural set up to explain such
neutrino masses through the seesaw 
mechanism~\cite{SeeSaw}\footnote{See Refs.~\cite{SeeSawanwendung} 
for early application of this mechanism.}. If
this is the case, the lepton number violation which is crucial for
leptogenesis is naturally explained. In fact, non-equilibrium decays
of Majorana neutrinos can produce a lepton number in the early
Universe, which is partially converted into a baryon number through
the electroweak sphaleron processes~\cite{Kuzmin:1985mm}. Therefore,
the Fukugita-Yanagida mechanism~\cite{Fukugita:1986hr}, called 
leptogenesis, is probably the most attractive possibility 
to generate dynamically the observed baryon asymmetry 
in the present Universe.

There is a variety of scenarios for leptogenesis in the
literature~\cite{Fukugita:1986hr,Buchmuller:2000as,Kumekawa:1994gx,Giudice:1999fb,Campbell:1992hd}.
Here we restrict ourselves to leptogenesis via decays of the seesaw
Majorana neutrinos ($N_i$: $i=1,2,3$) having the hierarchical masses 
$M_3\!\gg\!M_2\!\gg\!M_1$. It is usually considered that the decays of the
lightest Majorana neutrinos $N_1$ are responsible for the present
baryon asymmetry, although the decays of $N_2$ and $N_3$ also generate
a lepton asymmetry. This is because the $N_1$, having the lightest
mass, can remain in thermal equilibrium after the decays of $N_2$ and
$N_3$, and may induce additional rapid processes changing the lepton
number (other than the sphaleron process). These processes wash out
the lepton asymmetry from $N_2$ and $N_3$ ``too much''
to make $N_2$ and $N_3$ work as the producers of 
baryon number~\cite{Buchmuller:2002xm}, and hence the resultant asymmetry
only comes from the decays of $N_1$.

In this paper, however, we point out that the decays of the
{\em heavier} Majorana neutrinos $N_2$ and $N_3$ can be a dominant
source of the present baryon asymmetry if they are produced
{\em non-thermally} and also at {\em very late time}. The $N_2$ and $N_3$
produced non-thermally decay immediately after the production and
generate the lepton asymmetry through the Fukugita-Yanagida
mechanism. Furthermore, this production
time is so late that the cosmic temperature is low enough
to prevent thermalization of the $B-L$ (baryon minus lepton) 
asymmetry due to the 
lightest Majorana neutrinos $N_1$
as well as the wash-out processes caused by the existence of $N_1$.
As a result, after that time the $B-L$ quantum 
number becomes a good ``accidental symmetry'', which ensures
the sphaleron conversion from the lepton asymmetry from $N_2$ and $N_3$
into the baryon asymmetry.
This scheme is easily realized by a scalar field
which dominates the energy of the Universe
when it decays into the Majorana neutrinos with a long lifetime.
Suggestively this boson could be described by a scalar field which causes the
cosmic inflation, the inflaton, and thus its decay
means really the reheating of the Universe~\cite{Kumekawa:1994gx}.

The leptogenesis picture by the {\em heavier} Majorana neutrinos
produced non-thermally in inflaton decays works well
even if the leptogenesis by the $N_1$ decays is ineffective
and also even if the $N_1$-neutrinos are absent in the early Universe.
This means that our proposed scheme for the leptogenesis
can be applied to a wider class of models.
For example, we illustrate it in the family 
replicated gauge group model~\cite{NT,NT1} in which 
all the lepton and quark masses
and mixing angles are well fitted order of magnitudewise.
The proposed leptogenesis is well suited for this model, since
the leptogenesis by the decays of $N_1$ is not sufficient to explain
the phenomenological baryon asymmetry in this model. It is found 
that the proposed leptogenesis works well even in this
model and also, interestingly, we can determine
values of the reheating temperature as well as the inflaton mass
from the present baryon asymmetry using this 
family replicated gauge group model. 

The organization of this paper is as follows: In Section~\ref{sec2}
we briefly review the leptogenesis by using the inflaton decays
and we discuss the possibility of the baryon asymmetry generated
by the decays of the heavier Majorana neutrinos.
In Section~\ref{sec3} we review the family replicated gauge group model
in which we illustrate our idea. In Section~\ref{sec4} we then derive 
the $B-L$ asymmetry obtainable by the proposed ideas. Some 
problems of naturalness of the rather isolated inflaton or just 
scalar needed are discussed in Section~\ref{sec5} by suggesting 
some possible -- speculative -- solutions. Our conclusions 
go into Section~\ref{sec6}.

\section{Leptogenesis in the inflaton decays}
\label{sec2}

Let us start by explaining briefly the leptogenesis in 
inflaton decays scenario (see Ref.~\cite{Kumekawa:1994gx}).
After de Sitter expansion of inflation ends, the inflaton
decays, when the Hubble parameter of the Universe, $H$, becomes
comparable to the decay width of the inflaton $\Gamma_\phi$.
The vacuum energy of the inflaton $\phi$-field is completely released
into decay products, and the Universe is reheated through
their thermal scattering. The temperature at this time,
the reheating temperature, $T_R$, is given using $\Gamma_\phi$ by
\begin{eqnarray}
  \label{TR}
  T_R = 0.55 \sqrt{ M_* \Gamma_\phi}\nn,
\end{eqnarray}
where $M_* = 2.4 \times 10^{18}~\GeV$ is the reduced Planck mass.

We consider the seesaw Majorana neutrinos $N_i$ ($i=1,2,3$)
produced by the decays of the inflaton $\phi$.
If $N_i$ are produced ``{\it non-thermally}'',
the ratio between the number density of produced
$N_i$ and the entropy density $s$ is estimated to be
\begin{eqnarray}
  \frac{n_{N_i}}{s} = {\rm Br}_i \,\frac{3\, T_R}{ 2 \,m_\phi} \nn,
\end{eqnarray}
where ${\rm Br}_i$ denotes the branching ratio of the decay channel
$\phi \rightarrow N_i\, N_i$.
When the decay rate of $N_i$ is much larger than $\Gamma_\phi$,
the $N_i$ decays immediately after being produced by the 
inflaton decays. (As we shall discuss later, 
we are interested in low reheating temperatures
which ensure this condition.)

Decays of Majorana neutrinos $N_i$ break the lepton-number conservation
and have the $CP$ violation. There are two classes of decay channels:
\begin{eqnarray}
  N_i \rightarrow \phi_{\rm\sss WS} + \ell\nn, ~~~~
  N_i \rightarrow \phi_{\rm\sss WS}^\dagger + \overline{\ell}\nn,
\end{eqnarray}
where $\phi_{\rm\sss WS}$ and $\ell$ denote the Weinberg-Salam Higgs and
lepton doublets in the Standard Model, respectively.
The lepton asymmetry generated by $N_i$ decays can be expressed by
\begin{eqnarray}
\label{epM}
  \epsilon_i \equiv
  \frac{ \Gamma (N_i \rightarrow \phi_{\rm\sss WS} + \ell) -
         \Gamma (N_i \rightarrow \phi_{\rm\sss WS}^\dagger + 
\overline{\ell}) }{ \Gamma (N_i \rightarrow \phi_{\rm\sss WS} 
+ \ell) + \Gamma (N_i \rightarrow \phi_{\rm\sss WS}^\dagger
+ \overline{\ell}) }\nn.
\end{eqnarray}
Just after the reheating completes,
we obtain the lepton asymmetry induced by $N_i$ decays as
\begin{eqnarray}
  \label{LA}
\frac{ n_L}{s}
  = \sum_{i=1}^3\ \epsilon_i \, 
{\rm Br}_i \,\frac{3 \,T_R}{ 2 \,m_\phi}\nn.
\end{eqnarray}

We will assume for a while an accidental $B-L$ conservation (in the
sense that it is not imposed upon the model, but comes out). If the lepton 
asymmetry in (\ref{LA}) is produced well before
the electroweak phase transition of the thermal history, $\ie$,
$T_R \gg 100~\GeV$, the $B$ and $L$ conversion by the sphaleron process
is active, and it brings a part of this lepton asymmetry
into the baryon asymmetry as~\cite{Khlebnikov:sr}
\begin{eqnarray}
  \frac{n_B}{s} = -\,\frac{28}{79} \,\frac{n_L}{s}\nn.
\end{eqnarray}
Here we have assumed that there is only one weak Higgs 
doublet. Finally, 
we obtain the following expression of the produced baryon
asymmetry~\cite{Kumekawa:1994gx}
\begin{eqnarray}
  \label{EBA}
  \frac{n_B}{s}= -\,\frac{42}{79} \,\sum_{i=1}^3\,\epsilon_i \,{\rm Br}_i\, 
\frac{T_R}{m_\phi}\nn.
\end{eqnarray}

Now we are at the point of justifying the assumptions which we
have made in the above discussion.
First, the production of $N_i$ in the inflaton decays
is available only when
\begin{eqnarray}
  \label{C1}
  m_\phi>2\,M_i\nn.
\end{eqnarray}
Second, the estimation of the lepton asymmetry in Eq.~(\ref{LA}) 
is obtained under the requirement that the $N_i$ 
are produced non-thermally by the inflaton decays,
which leads to the following condition on the reheating temperature
\begin{eqnarray}
  \label{C2}
  T_R<\kappa_i \,M_i\nn,
\end{eqnarray}
where $\kappa_i$ are constants of order one defined by the 
decoupling temperatures
of $N_i$ to be $T_{i}^{dec} = \kappa_i M_i$.
Therefore, considering the inflation model satisfying the conditions
(\ref{C1}) and (\ref{C2}), the decays of the Majorana neutrinos $N_i$
generate the lepton asymmetry which is given in Eq.~(\ref{LA}) 
just after the reheating.

However, the lepton asymmetry might be washed out, after it is produced,
by the lepton-number violating processes. The most dangerous
ones are the processes mediated by $N_1$, since $N_1$ is the lightest
Majorana neutrino, so that it survives and still can be 
produced in the thermal bath after $N_2$ and $N_3$ have 
disappeared.
If those processes are well in thermal equilibrium
the produced lepton asymmetry is washed-out
strongly~\cite{Buchmuller:2002xm}.
To avoid these wash-out processes, we have to invoke
that the production of the Majorana neutrinos ($\ie$ lepton asymmetry)
takes place at a sufficiently late time so that the wash-out processes have 
already decoupled
and been ineffective. Thus, we have to consider
sufficiently low reheating temperatures,
\begin{eqnarray}
  \label{C3}
  T_R<\kappa_1 \,M_1\nn.
\end{eqnarray}
With such low reheating temperatures, the lighter Majorana neutrino(s)
are completely decoupled from the thermal bath of the Universe,
and the $B-L$ becomes a good ``accidental symmetry'' for $T<T_R$.
The conditions Eqs.~(\ref{C1}) and (\ref{C3}) justify
our estimation of the baryon asymmetry in Eq.~(\ref{EBA}), by 
explaining why there is no wash-out effect.

It should be noted that Eq.~(\ref{C3}) ensures the non-thermal
production of the heavier Majorana neutrinos $N_2$ and $N_3$
if $m_\phi>2M_3$. This means that the decays of $N_2$ and
$N_3$ can be dominant sources of the present baryon asymmetry
(if $\epsilon_2\,{\rm Br}_2$, 
$\epsilon_3\,{\rm Br}_3\gg\epsilon_1\,{\rm Br}_1$).
Further, this mechanism works even if there is no $N_1$, $\ie$, 
if ${\rm Br}_1=0$. This feature is completely different from the 
conventional thermal
leptogenesis~\cite{Buchmuller:2000as} where the lepton asymmetry 
is generated
by the decays of $N_2$ ($N_3$) at the temperature of
$T\sim M_2(M_3)\gg M_1$,
and hence the produced lepton asymmetry may be easily washed out
and the resultant baryon asymmetry comes from the decays of
the lightest Majorana neutrinos $N_1$.

These observations lead to that our proposed scheme for the leptogenesis
can be applied to a wider class of models of the sort explaining 
fermion masses and mixings.
In the next to next section, we will illustrate it in a specific example,
a model presented in the following section.

\section{Family replicated gauge group model}
\label{sec3}

In this section we review briefly the family replicated gauge
group model~\cite{NT,NT1}. This model
is based on a large gauge group being 
the Cartesian product of family specific gauge groups, namely,
\begin{equation}
  \label{eq:AGUTgauge}
  \AGUT \nn,
\end{equation}
where $SMG_i$ denotes $SU(3)_i\times SU(2)_i\times U(1)_i$ (Standard 
Model gauge group), $U_{\sss B-L,i}$ is $B-L$ gauge group, and $i$ 
denotes the generation. This 
group in Eq.~(\ref{eq:AGUTgauge}) is represented {\em only} 
by representations that do not mix 
the different irreducible representation of the Standard Model 
and it breaks spontaneously down to the 
group $SU(3)\times SU(2)\times U(1)\times U(1)_{\sss B-L}$
at the scale about $1$ to $2$ orders of magnitude under the Planck scale.
The breaking to the Standard Model gauge group is supposed to 
occur by means of five Higgs fields which we have 
invented and denoted by the symbols
$W$, $T$, $\rho$, $\omega$ and $\phi_{\rm\sss SS}$. Their quantum 
numbers are given in Table $1$. Finally the breaking of $SU(2)\times U(1)$
of the Standard Model is caused by a Weinberg-Salam Higgs field, 
$\phi_{\sss\rm WS}$ (also its quantum numbers are found 
in Table~$1$).

We summarize here the vacuum expectation values (VEVs) of the 
six Higgs fields which the model contains: %
\begin{list}{\it\arabic{line})}{\usecounter{line}}
\item The smallest VEV Higgs field is the
Weinberg-Salam Higgs field, $\phi_{\sss\rm WS}$,
with the VEV at the weak scale, $246~\GeV$.
\item The next smallest VEV Higgs field, called $\phi_{\sss\rm SS}$, 
is also alone in its class and breaks the ``diagonal $B-L$'' gauge group 
$U(1)_{\sss B-L}$, common to all the families. This symmetry is supposed 
to be broken (Higgsed) at the seesaw scale as needed for the 
neutrino oscillation scale in seesaw mechanism. This VEV turns out of the 
order of $10^{16}~\GeV$.
\item The next four Higgs fields are called $W$, $T$, $\rho$ and 
$\omega$ and have VEVs of the order of a 
factor $4$ to $50$ under the Planck unit. It means that if intermediate 
propagators have scales given by the Planck scale, as we assume, they 
will give rise to suppression factors of the order $1/4$ to $1/50$ each
time they are needed to cause a transition~\cite{FN}.
\end{list}

The quantum numbers of the $45$ well-known Weyl particles and 
the three additional Majorana neutrinos to be used as seesaw neutrinos  are 
strongly 
restricted by the requirement that all anomalies 
(gauge and mixed ones) shall vanish 
even without using Green-Schwarz anomaly cancellation 
mechanism~\cite{GS}. Thus the 
family replicated gauge group model is an anomaly free 
model. All the  $U(1)$ quantum charges  in this model can be found
in Table~$1$.


In the family replicated gauge group model there exist many bosons 
and fermions at the fundamental 
scale (the Planck scale).
However, in order to have no mass protection, the left-handed fermions 
and its right-handed partner must combine to a Dirac or 
vector-like-coupled fermion, in the sense that they are described as Dirac 
particles from the picture of the Weyl particles: they are 
combinations of left-handed and 
right-handed states with the same (gauge) quantum numbers. The left-over 
Weyl particles (in other word those without chiral partners) 
in this model -- $\ie$ those that could get small masses relative to the
Planck scale -- are specified in more detail and are actually assumed to 
form a system of three proto-generations, each consisting of the $16$ Weyl 
particles of a usual Standard Model generation plus one seesaw particle.
In this way we can label these particle as proto-left-handed or
proto-right-handed $u$-quark, $d$-quark, electron $\etc$. To get the 
quantum numbers in  the model under the  gauge group for a given fermion 
proto-irreducible representation, we proceed in the following way:
We note the generation number of the particle for which we want the 
quantum numbers and we look up, in the Standard Model, what the 
quantum numbers are for the irreducible representation in question 
and what the $B-L$ quantum number is. For instance, if we want to find 
the quantum numbers of the proto-right-handed strange quark, we note 
that the quantum numbers of the right-handed strange quark in the 
Standard Model are: weak hypercharge $y/2=-1/3$, 
singlet under $SU(2)$, and triplet under $SU(3)$, 
while $B-L$ is equal to the baryon number $=1/3$. Moreover, 
ignoring mixing angles (as we do for proto quarks), the 
generation is denoted as number $i=1$.
That it belongs to generation $i=1$ means that all the quantum numbers 
for $SMG_j$, $j=2,3$ are trivial. Also the baryon number minus 
lepton number for the
proto-generation number two and three are zero: only the quantum numbers 
associated with proto-generation one are non-trivial. Thus, in our model, 
the quantum numbers of the proto-right-handed down quark are 
$y_2/2 = -1/3$, singlet under $SU(2)_1$, triplet under 
$SU(3)_1$ and $(B-L)_1=1/3$. 

For each proto-generation the following charge 
quantization rule applies
\begin{equation}
  \label{eq:mod}
  \frac{t_i}{3}+\frac{d_i}{2} + \frac{y_i}{2} = 0~~{\rm (mod~1)}\nn,
\end{equation}
where $t_i$ and $d_i$ are the triality and duality for 
the $i$'th proto-generation gauge groups $SU(3)_i$ and $SU(2)_i$ 
respectively, of course a consequence of the corresponding rule in the
Standard Model (See Refs.~\cite{NT,NT1}).

Combining Eq.~(\ref{eq:mod}) with the principle of taking the smallest 
possible representation of the groups $SU(3)_i$ and $SU(2)_i$, 
is sufficient to specify the non abelian quantum numbers in terms of the 
six Abelian quantum numbers $y_i/2$ 
and $(B-L)_i$.
Using this rule we easily specify the fermion representations as in 
Table $1$.

\begin{table}[!t]
\caption{All $U(1)$ quantum charges in the family replicated model. 
The symbols for the fermions shall be considered to mean
``proto''-particles. Non-abelian representations are given by a rule 
from the abelian ones (see Eq.~(\ref{eq:mod})).}
\vspace{3mm}
\label{Table1}
\begin{center}
\begin{tabular}{|c||c|c|c|c|c|c|} \hline
& $SMG_1$& $SMG_2$ & $SMG_3$ & $U_{\sss B-L,1}$ & 
$U_{\sss B-L,2}$ & $U_{\sss B-L,3}$ \\ \hline\hline
$u_L,d_L$ &  $\frac{1}{6}$ & $0$ & $0$ & $\frac{1}{3}$ & $0$ & $0$ \\
$u_R$ &  $\frac{2}{3}$ & $0$ & $0$ & $\frac{1}{3}$ & $0$ & $0$ \\
$d_R$ & $-\frac{1}{3}$ & $0$ & $0$ & $\frac{1}{3}$ & $0$ & $0$ \\
$e_L, \nu_{e_{\sss L}}$ & $-\frac{1}{2}$ & $0$ & $0$ & $-1$ & $0$ 
& $0$ \\
$e_R$ & $-1$ & $0$ & $0$ & $-1$ & $0$ & $0$ \\
$\nu_{e_{\sss R}}$ &  $0$ & $0$ & $0$ & $-1$ & $0$ & $0$ \\ \hline
$c_L,s_L$ & $0$ & $\frac{1}{6}$ & $0$ & $0$ & $\frac{1}{3}$ & $0$ \\
$c_R$ &  $0$ & $\frac{2}{3}$ & $0$ & $0$ & $\frac{1}{3}$ & $0$ \\
$s_R$ & $0$ & $-\frac{1}{3}$ & $0$ & $0$ & $\frac{1}{3}$ & $0$\\
$\mu_L, \nu_{\mu_{\sss L}}$ & $0$ & $-\frac{1}{2}$ & $0$ & $0$ & 
$-1$ & $0$\\
$\mu_R$ & $0$ & $-1$ & $0$ & $0$  & $-1$ & $0$ \\
$\nu_{\mu_{\sss R}}$ &  $0$ & $0$ & $0$ & $0$ & $-1$ & $0$ \\ \hline
$t_L,b_L$ & $0$ & $0$ & $\frac{1}{6}$ & $0$ & $0$ & $\frac{1}{3}$ \\
$t_R$ &  $0$ & $0$ & $\frac{2}{3}$ & $0$ & $0$ & $\frac{1}{3}$ \\
$b_R$ & $0$ & $0$ & $-\frac{1}{3}$ & $0$ & $0$ & $\frac{1}{3}$\\
$\tau_L, \nu_{\tau_{\sss L}}$ & $0$ & $0$ & $-\frac{1}{2}$ & $0$ & 
$0$ & $-1$\\
$\tau_R$ & $0$ & $0$ & $-1$ & $0$ & $0$ & $-1$\\
$\nu_{\tau_{\sss R}}$ &  $0$ & $0$ & $0$ & $0$ & $0$ & $-1$ \\ 
\hline
$W$ & $0$ & $-\frac{1}{2}$ & $\frac{1}{2}$ & $0$ & $-\frac{1}{3}$ 
& $\frac{1}{3}$ \\ 
$T$ & $0$ & $-\frac{1}{6}$ & $\frac{1}{6}$ & $0$ & $0$ & $0$\\
$\rho$ & $0$ & $0$ & $0$ & $-\frac{1}{3}$ & $\frac{1}{3}$ & $0$\\
$\omega$ & $\frac{1}{6}$ & $-\frac{1}{6}$ & $0$ & $0$ & $0$ & $0$\\
$\phi_{\sss\rm SS}$ & $0$ & $1$ & $-1$ & $0$ & $2$ & $0$ \\ 
$\phi_{\sss\rm WS}$ & $0$ & $\frac{2}{3}$ & $-\frac{1}{6}$ & $0$ & 
$\frac{1}{3}$ & $-\frac{1}{3}$ \\
\hline
\end{tabular}
\end{center}
\end{table}

\subsection{Mass matrices}
\indent We can easily evaluate using the
system of quantum numbers given in Table~$1$ the numbers 
of Higgs field VEVs of the 
different types needed to construct the transition between 
the left- and right-handed Weyl fields corresponding to a 
given element in a mass matrix.

The main point is that the order 
of magnitudes of the mass matrix elements are 
determined by the number and order of magnitudes of the Higgs field
VEV factors needed to accomplish a given quantum number transition. 
We have namely assumed that all the 
Yukawa couplings, and also other couplings, as well as the masses of 
particles that are not
mass-protected are of order unity in some fundamental scale (which
we usually take to be the Planck scale). Thereby of course the whole 
calculation can only give order of magnitudewise results.

We shall now write down the mass matrices -- five Dirac type and one
Majorana neutrino mass matrix  -- which are necessary
to fit the fermion masses and mixing angles and also
to discuss the mechanism of the baryogenesis:

\noindent
the up-type quarks:
\begin{eqnarray}
M_{\sss U} \simeq \frac{\sVEV{(\phi_{\sss\rm WS})^\dagger}}{\sqrt{2}}
\hspace{-0.1cm}
\left(\!\begin{array}{ccc}
        (\omega^\dagger)^3 W^\dagger T^2
        & \omega \rho^\dagger W^\dagger T^2
        & \omega \rho^\dagger (W^\dagger)^2 T\\
        (\omega^\dagger)^4 \rho W^\dagger T^2
        &  W^\dagger T^2
        & (W^\dagger)^2 T\\
        (\omega^\dagger)^4 \rho
        & 1
        & W^\dagger T^\dagger
\end{array} \!\right)\label{M_U}
\end{eqnarray}  
\noindent
the down-type quarks:
\begin{eqnarray}
M_{\sss D} \simeq \frac{\sVEV{\phi_{\sss\rm WS}}}
{\sqrt{2}}\hspace{-0.1cm}
\left (\!\begin{array}{ccc}
        \omega^3 W (T^\dagger)^2
      & \omega \rho^\dagger W (T^\dagger)^2
      & \omega \rho^\dagger T^3 \\
        \omega^2 \rho W (T^\dagger)^2
      & W (T^\dagger)^2
      & T^3 \\
        \omega^2 \rho W^2 (T^\dagger)^4
      & W^2 (T^\dagger)^4
      & W T
                        \end{array} \!\right) \label{M_D}
\end{eqnarray}
\noindent %
the charged leptons:
\begin{eqnarray}        
M_{\sss E} \simeq \frac{\sVEV{\phi_{\sss\rm WS}}}
{\sqrt{2}}\hspace{-0.1cm}
\left(\hspace{-0.1 cm}\begin{array}{ccc}
    \omega^3 W (T^\dagger)^2
  & (\omega^\dagger)^3 \rho^3 W (T^\dagger)^2 
  & (\omega^\dagger)^3 \rho^3 W^4 (T^\dagger)^5\\
    \omega^6 (\rho^\dagger)^3  W (T^\dagger)^2 
  &   W (T^\dagger)^2 
  & W^4 (T^\dagger) ^5\\
    \omega^6 (\rho^\dagger)^3  (W^\dagger)^2 T^4 
  & (W^\dagger)^2 T^4
  & WT
\end{array} \hspace{-0.1cm}\right) \label{M_E}
\end{eqnarray}
\noindent
the Dirac neutrinos:
\begin{eqnarray}
M^D_\nu \simeq \frac{\sVEV{(\phi_{\sss\rm WS})^\dagger}}{\sqrt{2}}
\hspace{-0.1cm}
\left(\hspace{-0.1cm}\begin{array}{ccc}
        (\omega^\dagger)^3 W^\dagger T^2

        & (\omega^\dagger)^3 \rho^3 W^\dagger T^2
        & (\omega^\dagger)^3 \rho^3 W^2 (T^\dagger)^7\\ 
        (\rho^\dagger)^3 W^\dagger T^2
        &  W^\dagger T^2
        & W^2 (T ^\dagger)^7\\  
        (\rho^\dagger)^3 (W^\dagger)^4 T^8 
        & (W^\dagger)^4 T^8 
        & W^\dagger T^\dagger
\end{array} \hspace{-0.1 cm}\right)\label{Mdirac}
\end{eqnarray} 
\noindent %
and the Majorana (right-handed) neutrinos:
\begin{eqnarray}    
M_R \simeq \sVEV{\phi_{\sss\rm SS}}\hspace{-0.1cm}
\left (\hspace{-0.1 cm}\begin{array}{ccc}
(\rho^\dagger)^6 T^6 
& (\rho^\dagger)^3 T^6 
& (\rho^\dagger)^3 W^3 (T^\dagger)^3  \\
(\rho^\dagger)^3 T^6
& T^6 & W^3 (T^\dagger)^3 \\
(\rho^\dagger)^3 W^3 (T^\dagger)^3 & W^3 (T^\dagger)^3 & W^6 (T^\dagger)^{12}
\end{array} \hspace{-0.1 cm}\right ) \label{Majorana}
\end{eqnarray}       

We know neither the Yukawa couplings nor the precise 
masses of the fundamental fermions, but it is one of our basic 
assumptions of naturalness of the model that these 
couplings are of order unity and random complex
numbers at the Planck scale. In the numerical 
evaluation of the consequences of the model we explicitly take 
into account these uncertain factors of order unity by 
providing each matrix element with an explicit random number factor
$\lambda_{ij}$ with a distribution, so that its logarithm average 
$\sVEV{\log {\lambda_{ij}}}\approx 0$ and its spread is 
$64 \%$. At the end we then average the obtained results over these
random number distributions. Note that the random 
complex order of unity factors
which are supposed to multiply all the mass matrix elements are
not written explicitly in Eqs.~(\ref{M_U})-(\ref{Majorana})
but are understood to be there anyway.

The philosophy of the model is that these mass matrices correspond to 
effective Yukawa couplings to be identified with running Yukawa couplings 
at the fundamental/Planck scale for the Higgs field $\phi_{\sss\rm WS}$ 
in the case of the first three mass matrices and for 
$\phi_{\sss\rm SS}$ in the case of the right-handed neutrino 
mass matrix. Therefore these effective Yukawa couplings 
have in principle to be run down by the beta-functions to the scale of 
observation, see section~\ref{RGE}. It is also important that we include the 
``running'' of the irrelevant operator of dimension $5$ giving the neutrino
oscillation masses. 

The right-handed neutrino couplings -- or equivalently mass matrix elements 
-- are 
used to
produce an effective mass matrix for the left-handed neutrinos 
which we after the five dimensional running down mentioned take as the ones 
observed in neutrino oscillations~\cite{SeeSaw,SeeSawanwendung}:
\begin{equation}
M_{\rm eff} \simeq M_{\nu}^{D} M_R^{-1} (M_{\nu}^D)^T.
\end{equation}

\subsection{Renormalization group equations}
\label{RGE}
\indent

It should be kept in mind that the effective Yukawa couplings for 
the Weinberg-Salam Higgs field, which 
are given by the Higgs field factors in the above mass matrices 
multiplied by the understood order of unity factors (taken as random numbers), 
are the running (effective) Yukawa couplings at a scale {\em very 
close to the Planck scale}. Thus, in our calculations, we had to 
use the renormalization group 
$\beta$-functions to run these couplings down to the experimentally 
observable scale, $\ie$ $\mu=1~\GeV$ where $\mu$ is the renormalization 
point. This is because we 
took the charged fermion masses to be compared to ``measurements'' 
at the conventional scale of $1~\GeV$. In other words, what we 
take as input quark masses are the current algebra masses, 
corresponding to running masses at $1~\GeV$. We used though for the top quark 
the pole mass instead~\cite{pierre}:
\begin{equation}
M_t = m_t(M)\left(1+\frac{4}{3}\frac{\alpha_s(M)}{\pi}\right)\nn,
\end{equation}
where we set $M=180~\GeV$ as an input, for simplicity.

{}From the Planck scale down to the seesaw scale or rather from 
where our gauge group break down to $SMG\times U(1)_{B-L}$ we use
the one-loop renormalization group running of the Yukawa coupling constant 
matrices, $Y_{\sss U}$, $Y_{\sss D}$, $Y_{\sss E}$, $Y_{\sss \nu}$
and  $Y_{\sss R}$
(being proportional to the mass matrices $M_{\sss U}$, $M_{\sss D}$, 
$M_{\sss E}$, $M_{\sss\nu}^D$ and $M_{\sss R}$, respectively), and the 
gauge couplings~\cite{pierre,NT}:
\begin{eqnarray}
\label{eq:recha}
16 \pi^2 {d g_{1}\over d  t} &\!=\!& \frac{41}{10} \, g_1^3 \nn,\\
16 \pi^2 {d g_{2}\over d  t} &\!=\!& - \frac{19}{16} \, g_2^3 \nn, \\
16 \pi^2 {d g_{3}\over d  t} &\!=\!& - 7 \, g_3^3  \nn,\\
16 \pi^2 {d Y_{\sss U}\over d  t} &\!=\!& \frac{3}{2}\, 
\left( Y_{\sss U} (Y_{\sss U})^\dagger
-  Y_{\sss D} (Y_{\sss D})^\dagger\right)\, Y_{\sss U} 
+ \left\{\, Y_{\sss S} - \left(\frac{17}{20} g_1^2 
+ \frac{9}{4} g_2^2 + 8 g_3^2 \right) \right\}\, Y_{\sss U}\nn,\\
16 \pi^2 {d Y_{\sss D}\over d  t} &\!=\!& \frac{3}{2}\, 
\left( Y_{\sss D} (Y_{\sss D})^\dagger
-  Y_{\sss U} (Y_{\sss U})^\dagger\right)\,Y_{\sss D} 
+ \left\{\, Y_{\sss S} - \left(\frac{1}{4} g_1^2 
+ \frac{9}{4} g_2^2 + 8 g_3^2 \right) \right\}\, Y_{\sss D}\nn,\\
16 \pi^2 {d Y_{\sss E}\over d  t} &\!=\!& \frac{3}{2}\, 
\left( Y_{\sss E} (Y_{\sss E})^\dagger
-  Y_{\sss \nu} (Y_{\sss \nu})^\dagger\right)\,Y_{\sss E} 
+ \left\{\, Y_{\sss S} - \left(\frac{9}{4} g_1^2 
+ \frac{9}{4} g_2^2 \right) \right\}\, Y_{\sss E} \nn,\\
\label{Diracyukawa}
16 \pi^2 {d Y_{\sss \nu}\over d  t} &\!=\!& \frac{3}{2}\, 
\left( Y_{\sss \nu} (Y_{\sss \nu})^\dagger
-  Y_{\sss E} (Y_{\sss E})^\dagger\right)\,Y_{\sss \nu} 
+ \left\{\, Y_{\sss S} - \left(\frac{9}{20} g_1^2 
+ \frac{9}{4} g_2^2 \right) \right\}\, Y_{\sss \nu} \nn,\\
16 \pi^2 {d Y_{\sss R}\over d  t} &\!=\!& \left( (Y_{\sss \nu})^\dagger 
Y_{\sss \nu} \right)\,  Y_{\sss R} \,+\, Y_{\sss R} \,
\left( (Y_{\sss \nu})^\dagger Y_{\sss \nu} \right)^T\nn,\\
 \label{YScon} Y_{\sss S} &\!=\!& {\Tr}(\, 3\, Y_{\sss U}^\dagger\, 
Y_{\sss U} +  3\, Y_{\sss D}^\dagger \,Y_{\sss D} +  Y_{\sss E}^\dagger\, 
Y_{\sss E} +  Y_{\sss \nu}^\dagger\, Y_{\sss \nu}\,) \nn,
\end{eqnarray}
where $t=\ln\mu$ and $\mu$ is the renormalization point.

However, below the seesaw scale the right-handed neutrino
are no more relevant and the Dirac neutrino terms in the 
renormalization group equations should be removed, and also 
the Dirac neutrino Yukawa couplings themselves are not accessible
anymore. That means that, from the seesaw scale down to the 
weak scale, the only leptonic Yukawa $\beta$-functions
should become as follows:
\begin{equation}
16 \pi^2 {d Y_{\sss E}\over d  t} =\frac{3}{2}\, 
\left( Y_{\sss E} (Y_{\sss E})^\dagger \right)\,Y_{\sss E} 
+ \left\{\, Y_{\sss S} - \left(\frac{9}{4} g_1^2 
+ \frac{9}{4} g_2^2 \right) \right\}\, Y_{\sss E} \nn.
\end{equation}

Note that the quantity, $Y_{\sss S}$, must be also changed
below the seesaw scale:
\begin{equation}
\label{eq:Y_S}
Y_{\sss S}={\Tr}(\, 3\, Y_{\sss U}^\dagger\, Y_{\sss U} 
+  3\, Y_{\sss D}^\dagger \,Y_{\sss D} +  Y_{\sss E}^\dagger\, 
Y_{\sss E}\,)  \nn.
\end{equation}

In fact we stopped the running down according to formula 
(\ref{Diracyukawa}) differently for the different matrix 
elements in the $Y_\nu$ matrix corresponding to the right-handed 
neutrino mass supposed most important for the matrix element 
in question.

Starting the running in an analogous way, we further should evolve 
the effective neutrino mass matrix considered as two Higgs two fermion 
irrelevant -- a five dimensional -- term~\cite{5run} from the 
different right-handed 
neutrino masses to the weak scale ($180~\GeV$) depending on the terms:
\begin{equation}
\label{eq:remeff}
16 \pi^2 {d M_{\rm eff} \over d  t}
= ( - 3 g_2^2 + 2 \lambda + 2 Y_{\sss S} ) \,M_{\rm eff}
- {3\over 2} \left( M_{\rm eff}\, (Y_{\sss E} Y_{\sss E}^\dagger) 
+ (Y_{\sss E} Y_{\sss E}^\dagger)^T \,M_{\rm eff}\right) \nn,
\end{equation}
where $Y_{\sss S}$ defined in Eq.~(\ref{eq:Y_S}) and in this energy range
the Higgs self-coupling constant running equation is
\begin{equation}
\label{eq:rehiggs}
16 \pi^2 {d \lambda\over d  t}
= 12 \lambda^2 - \left( \frac{9}{5} g_1^2 + 9 g_2^2 \right) \,\lambda
+ \frac{9}{4} \left( \frac{3}{25} g_1^4 
+ \frac{2}{5} g_1^2 g_2^2 + g_2^4 \right) + 4 Y_{\sss S} \lambda 
- 4 H_{\sss S}\nn,
\end{equation}
with
\begin{equation}
 H_{\sss S} = {\Tr} \left\{ 3 \left(Y_{\sss U}^\dagger Y_{\sss U}\right)^2
 + 3 \left(Y_{\sss D}^\dagger Y_{\sss D}\right)^2 +  
\left(Y_{\sss E}^\dagger Y_{\sss E}\right)^2\right\} \nn.
\end{equation}
The mass of the Standard Model Higgs boson is given 
by $M_H^2 = \lambda \sVEV{\phi_{WS}}^2$ and, for definiteness, we 
took $M_H = 115~\GeV$ at weak scale.

{}From $180~\GeV$ down to $1~\GeV$ -- experimental 
scale ($1~\GeV$) -- we have evaluated the beta-functions with {\rm only} 
the gauge
coupling constants. In order to run the renormalization group
equations, we use the following initial values:
\begin{eqnarray}
U(1):\quad & g_1(M_Z) = 0.462 \nn,\quad & g_1(M_{\rm Planck}) = 0.614  \nn,\\
SU(2):\quad & g_2(M_Z) = 0.651 \nn,\quad & g_2(M_{\rm Planck}) = 0.504 \nn,\\
SU(3):\quad & g_3(M_Z) = 1.22  \nn,\quad & g_3(M_{\rm Planck}) = 0.491 \nn.
\end{eqnarray}
Note that we have ignored the influence of the 
$B-L$ gauge coupling constants; however, this effect should not be 
significant, because there is from the Planck scale to the seesaw scale 
$(\approx 10^{16}~\GeV)$ only a factor 
$10^3$. So it should be good
enough for our order magnitude calculations.

\subsection{$CP$ violation in Majorana sector}
\indent

The $CP$ violation in the Majorana neutrino decays in 
Eq.~(\ref{epM}), 
$\epsilon_i$, arises when the effects of loops are 
taken into account, and at the one-loop level, the $CP$
asymmetry comes both from the wave function renormalization 
and from the vertex correction~\cite{LCRVBP}:
\begin{equation}
\label{eq:CPepsilon}
\epsilon_i = \frac{1}{4 \pi \sVEV{\phi_{\sss WS}}^2 
((\Tilde{M_\nu^D})^{\dagger}\Tilde{M_\nu^D})_{ii}}\sum_{j\not= i} 
{\rm Im}[((\Tilde{M_\nu^D})^{\dagger} \Tilde{M_\nu^D})^2_{ji}] 
\left[\, f \left( \frac{M_j^2}{M_i^2} \right) 
+ g \left( \frac{M_j^2}{M_i^2} \right)\,\right] \nn,
\end{equation}
where $\Tilde{M_\nu^D}$ can be expressed through the 
unitary matrix diagonalizing of the right-handed neutrino 
mass matrix $V_R$:
\begin{eqnarray}
  \label{eq:tildemd}
 \Tilde{M_\nu^D} \!&\equiv&\! M_\nu^D\,V_R \nn,\\
V_R^\dagger \,M_R\,M_R^\dagger\, V_R \!&=&\! 
{\rm diag} \left(\,M^2_1, M^2_2, M^2_3\,\right) \nn.
\end{eqnarray}

The functions in Eq.~(\ref{eq:CPepsilon}), $f(x)$ and $g(x)$,
are obtained by the calculations in perturbation theory,
the one-loop vertex contribution and the self-energy contribution,
respectively, under the condition that the differences between Majorana
neutrino masses are sufficiently large compared to their
decay widths. Their precise
forms are as follows:
\begin{equation}
f(x) = \sqrt{x} \left[1-(1+x) \ln \frac{1+x}{x}\right]\nn,
\nn g(x) = \frac{\sqrt{x}}{1-x}\nn.
\end{equation}

We have taken into account the
renormalization group running effects in all sectors,
not only the charged sectors, but also Dirac- and
Majorana neutrino sectors to evaluate from
the Planck scale down to
the scale, $1~\GeV$. Furthermore, we have even taken into account
the running of the
dimension five operator involving two $\phi_{\rm\sss WS}$'s
and two left-handed neutrinos which give the neutrino
oscillation masses. The $\epsilon_i$'s were obtained
at the corresponding temperature, $\ie$, the Majorana neutrino masses
$(T=M_i)$.

\subsection{Mass and mixing results}

\begin{center}
\begin{table}[!t]
\caption{Best fit to conventional experimental data.
All masses are running
masses at $1~\GeV$ except the top quark mass which is the pole mass.
Note that we use the square roots of the neutrino data in this
Table, as the fitted neutrino mass and mixing parameters,
in our goodness of fit ($\mbox{\rm g.o.f.}$) definition,
Eq.~(\ref{gof}).}
\begin{displaymath}
\begin{array}{|c|c|c||c|c|c|}
\hline
 & {\rm Fitted} & {\rm Experiment} &  & {\rm Fitted} & {\rm Experiment} \\
 \hline
m_u & 4.9~\MeV & 4~\MeV
& m_d & 4.8~\MeV & 9~\MeV \\
m_c & 0.68~\GeV & 1.4~\GeV
& m_s & 325~\MeV & 200~\MeV \\
M_t & 202~\GeV & 180~\GeV
& m_b & 6.0~\GeV & 6.3~\GeV \\
m_e & 1.7~\MeV & 0.5~\MeV
& V_{us} & 0.11 & 0.22 \\
m_{\mu} & 122~\MeV & 105~\MeV
& V_{cb} & 0.028 & 0.041 \\
m_{\tau} & 1.53~\GeV & 1.78~\GeV
& V_{ub} & 0.0030 & 0.0035 \\ \hline
\tan^2\theta_{\odot} & 0.27 & 0.38 &
\Delta m^2_{\odot} & 9.0 \times 10^{-5}~\eV^2 &  6.9 \times 10^{-5}~\eV^2 \\
\tan^2\theta_{\rm atm}& 0.55 & 1.0 & \Delta m^2_{\rm atm} & 1.8\times 10^{-3}~\eV^2 &  2.5\times 10^{-3}~\eV^2 \\ \cline{4-6}
\tan^2\theta_{\rm chooz}  & 2.57 \times 10^{-2} & \sleq~2.6 \times 10^{-2} &
\mbox{\rm g.o.f.} &  3.51 & - \\
\hline\hline
& {\rm Prediction} & {\rm Experiment} & & {\rm Prediction} & {\rm Experiment}
\\ \hline
J_{\sss CP} & 4.4\times10^{-6} & (2-3.5)\times10^{-5} &
\sVEV{\abs{m_{ee}}} & 3.3\times10^{-3}~\eV & (0.11-0.56)~\eV\\
\tau{\sss(p\to\pi^0 e^+)} & \sgeq~10^{43}~{\rm yrs} & \sgeq~2.6\times10^{33}~{\rm yrs}&
{\rm Br}{\sss(\mu\to e\gamma)}& \sim~7.0\times10^{-56} & \sleq~1.2\times10^{-11} \\ \hline
\end{array}
\end{displaymath}
\label{convbestfit}
\end{table}
\end{center}
The calculation using random numbers and performed numerically
was used to fit the masses and mixing angles to the phenomenological
estimates by minimizing what we call ``goodness of fit'',
\begin{equation}
\label{gof}
\mbox{\rm g.o.f.}\equiv\sum_i \left[\ln\left(
\frac{m_{i \,, {\rm fit}}}{m_{i\,, {\rm exp}}}\right) \right]^2 \nn,
\end{equation}
a kind of $\chi^2$ for the case that we have only order of
magnitude accuracy. The results presented in Table~$2$ were obtained
from the following values of the
set of Higgs VEVs -- where the Higgs field VEVs for the fields
$W$, $T$, $\rho$ and $\omega$ causing the breaking to the diagonal subgroup
$SMG\times U(1)_{B-L}$ are quoted with the VEV in Planck units,
while they are for $\phi_{\sss\rm SS}$ (and the not fitted
$\phi_{\sss\rm WS}$) given in $\GeV$ units:
\begin{eqnarray}
\label{bestvevs}
&&\sVEV{W}=0.157\nn,\nn \sVEV{T}=0.0804\nn,
\nn\sVEV{\rho}=0.265\nn, \nn\sVEV{\omega}=0.244\nonumber\\
&&\sVEV{\phi_{\sss\rm SS}}=5.25\times10^{15}~\GeV\nn,
\nn\sVEV{\phi_{\sss\rm WS}}=246~\GeV\nn.
\end{eqnarray}
The results of the best fit, with the VEVs in Eq.~(\ref{bestvevs}),
are shown in Table~$2$ and the fit has
$\mbox{\rm g.o.f.}=3.51$. To see a typical error, say average
error, compared to the experimental values we should divide
this value with the number of predictions $(17-5=12)$ and then
take the square root of it: $\sqrt{3.51/12}=0.54$. This means
that the $12$ degrees of freedom have each of
them a logarithmic deviation of about $55\%$, $\ie$, we have
fitted {\rm all quantities} with a typical error
of a factor $\exp\left(\pm\sqrt{3.51/12}\right)\simeq{}1.00^{+0.72}_{-0.58}$.
This agrees with theoretically predicted deviations~\cite{FF}.

Note that to make the best fit values of the mass squared difference 
and mixing angle parameters presented in Table~$2$ 
we used the following point 
as fit values from the combination of the 
latest Super-Kamiokande and SNO experimental 
results~\cite{SNOuSK}:
\begin{equation}
 \Delta m^2_\odot=6.9\times 10^{-5}~\eV^2 
~~{\rm and}~~\tan^2\theta_{\odot}=0.38\nn. 
\end{equation}
The atmospheric neutrino parameters are the following according to
the Super-Kamio\-kande results~\cite{SKatm}:
\begin{equation}
 \Delta m^2_{\rm atm}=2.5\times 10^{-3}~\eV^2 
~~{\rm and}~~\tan^2\theta_{\rm atm}=1.0\nn. 
\end{equation}

With the VEVs in Eq.~(\ref{bestvevs}) we can calculate several
physical observables~\cite{Groom:in} -- Jarlskog 
triangle area~\cite{Jarlskog}
$J_{\sss CP}$,
proton lifetime $\tau{(p\to\pi^0 e^+)}$, the effective
electro-neutrino mass $\sVEV{\abs{m_{ee}}}$ relevant for the neutrinoless
double beta decay~\cite{klapdor}, and the branching
ratio ${\rm Br}(\mu\to e\gamma)$ --
which can be found also in Table~$2$.

We also present our predicted hierarchical left-handed neutrino masses
$(m_i)$ and the right-handed neutrino masses $(M_i)$
as well as the $CP$ violation in the Majorana neutrino decays
$(\epsilon_i)$ with mass eigenstate indices $(i=1,2,3)$:
\begin{eqnarray}
\label{kekka}
m_1 &\!=\!& 1.3\times 10^{-3}~~\eV\nn,\nn\nn M_1 = 1.4\times10^{6}~\GeV\nn,
\nn\nn\abs{\epsilon_1}=1.8 \times 10^{-10}\nn,\nonumber\\
m_2 &\!=\!& 9.6\times 10^{-3}~~\eV\nn,\nn\nn M_2 = 1.4\times10^{10}~\GeV\nn,\nn\nn\abs{\epsilon_2} = 2.2 \times 10^{-6}\nn, \\
m_3 &\!=\!& 4.3\times 10^{-2}~~\eV\nn,\nn\nn M_3 = 1.6\times10^{10}~\GeV\nn,\nn\nn\abs{\epsilon_3} = 1.8 \times 10^{-6}\nn.\nonumber
\end{eqnarray}

Due to the philosophy that all coupling constants are of order unity
and {\em complex random numbers} at fundamental scale, $\ie$,
the phases are rotated randomly. So we are not able to
predict the sign of these quantities. Therefore, we present
absolute values for the $CP$-violating parameters in the Majorana sector.

We estimate for later calculations of baryon asymmetry
in this model the uncertainties for the calculated
physical quantities. The order unity complex numbers are given
by a Gaussian distribution with mean value zero, which leads to a fluctuation 
in the logarithm of $\pm 64\%$. So the expected fluctuation 
in the natural exponent
is $64\%\cdot\sqrt{8}$ for the $\epsilon$'s (straight forward
assumption of independence of ${\cal O}(1)$ factor fluctuations
in numerator and denominator of Eq.~(\ref{eq:CPepsilon})
would lead to $64\%\cdot\sqrt{11}$, but crude expectation
of compensating correlations may justify
$64\%\cdot\sqrt{8}$) and $64\%$
for the right-handed neutrino masses, respectively. Here are 
these quantities 
with the errors:
\begin{eqnarray}
\label{errors}
&& M_1 = 1.4^{+1.3}_{-0.66}\times10^{6}~\GeV\nn,
\nn\nn\abs{\epsilon_1}=1.8^{+9.2}_{-1.5}\times 10^{-10}\nn,\nonumber\\*[2mm]
&& M_2 = 1.4^{+1.3}_{-0.66}\times10^{10}~\GeV\nn,
\nn\nn\abs{\epsilon_2} = 2.2^{+11}_{-1.8}\times 10^{-6}\nn, \\*[2mm]
&& M_3 = 1.6^{+1.4}_{-0.76}\times10^{10}~\GeV\nn,
\nn\nn\abs{\epsilon_3} = 1.8^{+9.2}_{-1.5}\times 10^{-6}\nn.\nonumber
\end{eqnarray}

\section{Estimation of $B-L$ production}
\label{sec4}

We shall discuss here the $B-L$ production in 
the model described above.
It is found from the results given in Eq.~(\ref{kekka}) that
the leptogenesis induced by the lightest Majorana neutrinos $N_1$ 
cannot account for the present baryon asymmetry (see Eq.~(\ref{BA})), 
since the decay rate asymmetry $\abs{\epsilon_1}$ 
for $N_1$ is too small. Therefore, we have to investigate the possibility 
of the leptogenesis by the {\em heavier} ones, $N_2$ and $N_3$,
being then produced non-thermally in inflaton decays.

With well-suited assumptions on the inflaton mass $m_\phi$ and the
reheating temperature $T_R$ the formula (\ref{EBA}) gives the baryon
asymmetry produced by decays of $N_2$ and $N_3$ as
\begin{eqnarray}
  \frac{n_B}{s} \!&\simeq&\! \frac{42}{79} \,\sum_{i=2}^3\,\abs{\epsilon_i} 
\,{\rm Br}_i\, \frac{T_R}{m_\phi} \label{modbaryo}\\
\!&\approx&\!  \left( 1.2 \times 10^{-6} \,{\rm Br}_2 + 0.96 \times 10^{-6} 
\,{\rm Br}_3 \right)\,\frac{ T_R }{ m_\phi }\nn.
\end{eqnarray}
As we have already mentioned that the applied model is not able to predict 
the sign of the considered quantities, $\epsilon$'s, we will use 
Eq.~(\ref{modbaryo}) instead of Eq.~(\ref{EBA})  
for baryogenesis calculation.
Here and hereafter we use the central values of physical quantities
given in Eq.~(\ref{kekka}) in the calculation.

First of all, we make the rough estimate on the possible value
of the baryon asymmetry generated by the proposed mechanism.
The condition on $m_\phi$ in Eq.~(\ref{C1}) leads to
\begin{equation}
  \frac{n_B}{s} \,\sleq \,\left( 1.2 \times 10^{-6} \,{\rm Br}_2 
+ 0.96 \times 10^{-6} \,{\rm Br}_3 \right)\frac{ T_R }{ 2 M_3 }\nn.
\end{equation}
Moreover, the reheating temperature should be low enough to
avoid the wash-out processes by $N_1$ as explained in Eq.~(\ref{C3}).
By taking $\kappa_1=1$, we find the produced baryon asymmetry to be
\begin{eqnarray}
\frac{n_B}{s} \!&\sleq&\!
  \left( 1.2 \times 10^{-6} \,{\rm Br}_2 + 0.96 \times 10^{-6} \,
{\rm Br}_3 \right) \,\frac{ M_1 }{ 2 M_3 }\nonumber \\
  \!&=&\! 5.1\times 10^{-11} \,{\rm Br}_2 + 4.2\times 10^{-11} 
\,{\rm Br}_3 \nn,
\end{eqnarray}
which gives, in the extreme case of ${\rm Br}_2={\rm Br}_3=1/2$,
\begin{equation}
  \frac{n_B}{s} \,\sleq\, 4.7 \times 10^{-11}\nn.
\end{equation}
Therefore, broadly speaking, in the family replicated gauge group model 
described in the previous section
the observed baryon asymmetry (\ref{BA}) can be explained by 
invoking the proposed non-thermal leptogenesis.

This non-thermal scenario for the leptogenesis requires the reheating 
temperature in the region of
\begin{eqnarray}
  \label{TRM1}
  T_R ~\sleq~ \kappa_1\,M_1 \sim 10^6~\GeV\nn.
\end{eqnarray}
It is important to mention that such low reheating temperatures
can be naturally obtained in the considered model.
We expect naturally that the inflaton $\phi$ is totally singlet 
under the gauge groups of the model given in Eq.~(\ref{eq:AGUTgauge})
in order to prevent the large radiative corrections disturbing the flatness 
of the inflaton potential. Then, we obtain the couplings 
between inflaton $\phi$ and Majorana neutrinos $N_i$ as
\begin{eqnarray}
  \label{Lint}
  {\cal L} &=& \frac{\phi}{M_*} \left( \mbox{gauge invariant terms} \right)
  \nonumber \\
  &\supset&
   \frac{\phi}{M_*}\,
   c_i\,M_i~ N_i\, N_i\nn,
\end{eqnarray}
where we have taken the effective cut-off scale as the reduced Plank scale
$M_*$. It is crucial to notice that the strength of these interactions
are determined up to ${\cal O}(1)$ factors $c_i$
by the charges of the Majorana neutrinos 
under the flavor (gauge) symmetries~\cite{FN}.
The inflaton might couple to other fermions and/or Higgs fields
similar to $N_i$, which might disturb our leptogenesis
by raising the reheating temperature and by lowering the branching
ratios ${\rm Br}_i$ (See the discussion in the next section.).

It is found from Eq.~(\ref{Lint}) that the dominant channels of the
inflaton decay are $\phi\rightarrow N_2\,N_2$ and/or $\phi\rightarrow
N_3\,N_3$ when $m_\phi >2 M_3$, and the total decay rate of the inflaton
is given by
\begin{equation}
  \label{GAM_P}
  \Gamma_\phi \simeq \frac{1}{4\pi} \,m_\phi\,
  \sum_{i=2}^3 \,\abs{c_i}^2\,\left( \frac{ M_i }{ M_* } \right)^2\,
  \left( 1 - \frac{ 4 M_i^2 }{ m_\phi^2 } \right)^{3/2}\nn.
\end{equation}
The reheating temperature is estimated from Eq.~(\ref{TR}),
by neglecting the factor \mbox{$( 1 - 4 M_i^2/m_\phi^2 )^{3/2}$}
and by taking $\abs{c_2}=\abs{c_3}=1$, as
\begin{eqnarray}
  T_R
  &\simeq&
  0.16 ~ 
  \sqrt{  \left( \frac{M_2}{M_\ast} \right)^2
    + \left( \frac{M_3}{M_\ast} \right)^2 }
  ~\sqrt{ m_\phi M_\ast } ~
  \nonumber \\
  &=&
  3.1 \times 10^{5}~\GeV
  \times
  \left( \frac{ m_\phi }{ 10^{11}~\GeV } \right)^{1/2}\nn.
\end{eqnarray}
Interestingly, it shows that the required reheating temperatures in
Eq.~(\ref{TRM1}) are naturally obtained when $m_\phi \sim 2 M_{2,3}$
with couplings $\abs{c_{2,3}}$ of order one.
Therefore, the interaction Lagrangian (\ref{Lint}) ensures the late decay
of the inflaton, $\ie$, the non-thermal leptogenesis at late time.

\begin{figure}[t!]
    \centerline{\psfig{figure=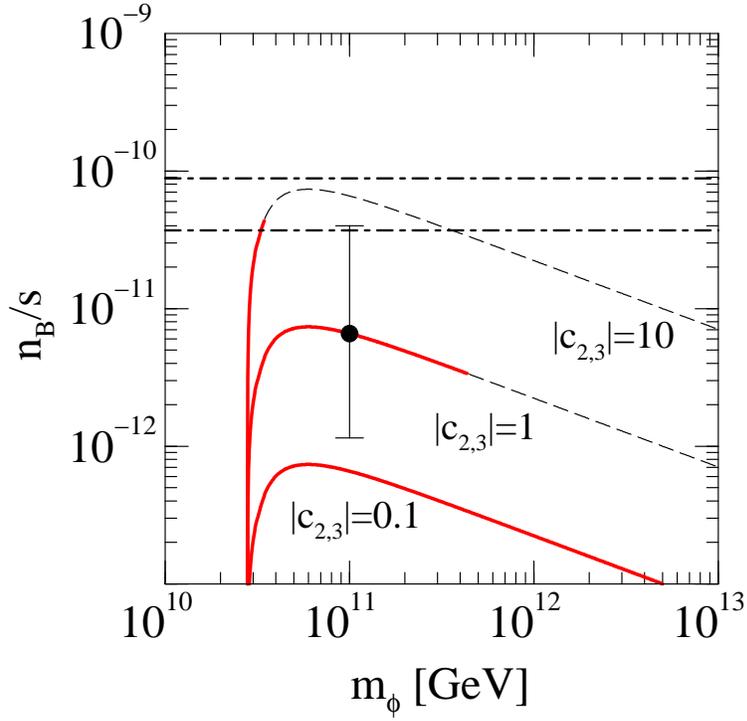,width=10cm}}
    \caption{
      The produced baryon asymmetries $n_B/s$ are shown by
      the (red) thick-solid lines for $\abs{c_{2,3}}=0.1, 1$ and $10$, 
      respectively. The dashed lines represent the regions 
      where the reheating temperature is higher than the lightest Majorana
      mass $T_R>M_1$.
      The typical effect of the statistical error of the $CP$ asymmetries 
      $\epsilon_{2,3}$ is also represented for $m_\phi=1\times10^{11}~\GeV$ 
      and $\abs{c_{2,3}} = 1$.
      The observed $n_B/s$ are denoted by the region 
      between the horizontal dot-dashed lines.
    }
    {\protect \label{fig:fig}}
\end{figure}

In Figure~\ref{fig:fig} we show the produced baryon asymmetries 
(\ref{modbaryo}) by using the exact formula for the inflaton decay 
width in Eq.~(\ref{GAM_P}). We neglect the region where the reheating 
temperature is higher than the lightest Majorana neutrino mass
$T_R>M_1$. It can be seen that the observed baryon asymmetry 
is obtained when the inflaton mass is close to $2 M_{2,3}$ 
with $\abs{c_{2,3}} \sim 10$.
The inflaton mass of $m_\phi\sim2M_{2,3}$ but $m_\phi>2M_{2,3}$
is crucial to have a sufficiently low reheating temperature 
$T_R\,\sleq\, M_1$ in order to avoid the wash-out of the produced asymmetry 
by $N_1$, as well as to have a larger baryon asymmetry.

One might worry that we only get agreement with the 
baryon asymmetry for the value $10$ of the order one constants,
$c_2=c_3=10$, given in Eq.~(\ref{Lint}). However, it is basically not a
problem at all: we have used the best fitted values for the  
$CP$-asymmetry parameters $\epsilon_{2,3}$, but they have large statistical 
uncertainty as represented in Eq.~(\ref{errors}). We show this typical error
also in Figure~\ref{fig:fig}. It is found that the observed baryon asymmetry
is easily obtained with $\abs{c_{2,3}} \sim$ a few, which is very consistent
with the philosophy that every dimensionless coupling takes an order one 
value at the fundamental scale. 

To summarize, we have found that the proposed non-thermal scenario 
for the leptogenesis indeed works well in the family replicated gauge 
group model, and the dominant contributions to the present baryon asymmetry
comes from the out-of-equilibrium decays of the heavier Majorana neutrinos
$N_2$ and $N_3$ rather than the lightest one, $N_1$.
To avoid the strong wash-out of the produced lepton asymmetry mediated 
by $N_1$, it is crucial that the inflaton decays 
into $N_2$ and $N_3$ at the late time
via the interaction terms in Eq.~(\ref{Lint}), which 
ensures naturally the reheating temperature of sufficiently low.
Interesting enough, we have observed that the successful baryogenesis 
requires (or predicts) the mass of the inflaton of 
$m_{\phi} \sim 2 M_{2,3} \sim 10^{10}~\GeV$.

\section{Discussion} 
\label{sec5}
\subsection{Problem of decay into Weinberg-Salam Higgs particles}
{}For the successful baryogenesis the inflaton particle should survive
until the era of the temperature $T\sleq M_1$, when the ``accidental''
conservation law of the $B-L$ charge has been installed to prevent the
wash-out of the produced lepton asymmetry. Furthermore, 
the branching ratio of the inflaton decays 
into the heavier Majorana neutrinos $N_2$ and $N_3$ should 
be sufficiently large to account for the observed
baryon asymmetry. To realize this long-lived
inflaton it is crucial that the total decay width of the inflaton
$\Gamma_\phi$ is really dominated by the decays into the Majorana
neutrinos $N_2$ and $N_3$ via the interaction (\ref{Lint}).

In the family replicated gauge group model discussed above, other
inflaton decay channels than the Majorana neutrinos do exist. In
fact, the inflaton can also decay into pairs of quarks or charged
leptons. These decay amplitudes are proportional to the masses of the
produced fermions, and hence it is naturally understandable that the
heaviest fermions ${N_2}$ and $N_3$ will dominate among the final
states of two fermions. Moreover, by assuming the philosophy that the
dimensionless couplings are of order one, we naturally expect that the
Higgs fields in the considered model obtain the masses of order of
their VEVs. Among these the Higgs fields, the masses of $W$, $T$,
$\rho$, $\omega$, and $\phi_{\sss\rm SS}$ are sufficiently heavy (see
Eq.~(\ref{bestvevs})) that the inflaton of mass $m_\phi \sim 2
M_{2,3}$ cannot kinematically decay into them.  The inflaton also
decays into pairs of gauge bosons in the Standard Model, \ie, photons,
$W^\pm$-bosons, $Z^0$-bosons, and gluons.  These decay channels would
be potentially dangerous depending on the inflaton mass.  However, in
the inflaton mass region we are interested in (\ie, $m_{\phi} \sim 2
M_{2,3}$ but well above threshold) the partial widths of decays into
the Standard Model gauge bosons are comparable to those of $\phi
\rightarrow N_{2,3} N_{2,3}$.  Therefore, they do not disturb our
successful leptogenesis scenario.  Other gauge fields appear in the
considered model obtain the masses from the Higgs fields $W$, $T$,
$\rho$, $\omega$, and $\phi_{\sss\rm SS}$, and the inflaton decays
into these gauge fields are kinematically forbidden.

However, there is one point that needs discussions:
why should the inflaton {\em not decay} into a couple of
Weinberg-Salam Higgs bosons, one Higgs and one anti-Higgs?
{\it A priori} there could exist interaction terms in the Lagrangian
density of the form
\begin{equation}
  \label{Higgscouplings}
  \cL=  \mu ~\phi ~ \abs{\phi_{\rm\sss WS}}^2\,
  +
  \lambda\, \phi^2 \, \abs{\phi_{\rm\sss WS}}^2\,,
\end{equation}
where $\mu$ is a mass parameter and $\lambda$
a coupling constant.  These interactions allow the inflaton decay with a width
\begin{eqnarray}
\label{GAM_H}
  \Gamma(\phi\rightarrow\phi_{\rm\sss WS}\,
\phi_{\rm\sss WS}^\dagger) = \frac{1}{ 8 \pi } 
  \frac{ \left( \mu + \lambda \sVEV{\phi} \right)^2 }{ m_\phi }~.
\end{eqnarray}
That would, however, have the adverse
effect of allowing the fast decay of the inflaton and making the
formal reheating temperature $T_R$ in Eq.~(\ref{TR}) higher than the
mass of $\phi$, $m_\phi$, since we expect from our philosophy 
the coupling $\lambda$ be the order unity and also 
the inflaton and its VEV be of the same order, $m_\phi \sim \sVEV{\phi}$.
It means that the $\phi$ inflaton would decay
immediately after the inflation, with exceedingly small branching
rations to seesaw neutrinos and our picture would be spoiled. 
In practice, in order that
the $\phi_{\rm\sss WS}\,\phi_{\rm\sss WS}^\dagger$-channel leave the
Majorana neutrino channel(s) to dominate, the required suppression 
of the effective coupling $\lambda_{\rm\sss eff}$
compared to unity is found from
Eqs.~(\ref{GAM_P}) and (\ref{GAM_H}) to be of the order of
\begin{eqnarray}
  \lambda_{\rm\sss eff} = \lambda + \frac{\mu}{\sVEV{\phi}} &\ll& 
  \abs{c_{2,3}} \frac{m_\phi}{\sVEV{\phi}}\,\frac{M_{2,3}}{M_\ast} 
  \nonumber\\
  &\approx& \abs{c_{2,3}} \frac{M_{2,3}}{M_\ast} 
  \approx 10^{-8} \cdot \abs{c_{2,3}}
  \left(\frac{M_{2,3}}{10^{10}~\GeV}\right)\nn,
  \label{suppression}
\end{eqnarray}
where we assumed that the mass of the inflaton and its VEV are
of the same order, $m_\phi/\sVEV{\phi}\approx1$.
Therefore, to realize the successful leptogenesis the coefficients
$\lambda$ and $\mu/\sVEV{\phi}$ should be extremely small
contrary to our basic philosophy.
This is a drawback of the proposed scenario.

However, first of all, we must say that we had already accepted this
kind of finetuning of parameters in the model. Since we have no
symmetry to control radiative corrections, we would expect every mass
parameter in the model to be of the order of the fundamental cutoff
scale, $\eg$, the Planck mass, under the assumption of dimensionless
coupling being order unity. On the contrary, we had already
introduced the mass scales which is smaller than the Planck scale,
which are the VEVs (or masses) of the Higgs fields.  The most severe
one is the mass of the Weinberg-Salam Higgs.  Although the hierarchy
in various Yukawa couplings to explain the observed mass spectrum of
fermions can be very elegantly explained by the flavor (gauge) 
symmetry, we had
not explained the hierarchy between the Planck scale and the mass
scales of interest, but just realized it by finetuning by hand.  In
fact, as an example, the order unity coupling 
$\lambda_{\rm\sss eff}={\cal O}(1)$ also induces the huge 
correction to the mass of the Weinberg-Salam Higgs particle, $\ie$, 
$\delta m_{\rm\sss WS} =\sqrt{ \lambda_{\rm\sss eff}} \sVEV{\phi}
\sim\sqrt{\lambda_{\rm\sss eff} }\, m_\phi\sim10^{10}~\GeV$. To 
have the weak scale mass
$\sim 10^2~\GeV$ for the Weinberg-Salam Higgs, the coupling would be
extremely small $\lambda_{\rm\sss eff} \ll 10^{-16}$, which is more
stringent than Eq.~(\ref{suppression}). However, this is not a matter
at all from our standpoint, since, in addition to other dangerous
(radiative) corrections, such a huge contribution is tremendously
tuned so as to have the Weinberg-Salam Higgs mass of the weak scale
$\sim 100~\GeV$. In the considered model we had already accepted the
finetuning for the mass parameters in the scalar sector. Therefore,
the required suppression of the coefficients as given
in Eq.~(\ref{suppression}), which makes the non-thermal leptogenesis works
well, is a problem of our attitude toward the finetuning.
Although we loose the simple philosophy of couplings being order
unity, the successful leptogenesis might be obtained just by accepting
not only the finetuning for the hierarchy problem, but also of the
dimensionless couplings appearing in the scalar sector. Moreover, it
should be remarked that inflaton scheme in general suffers from
difficulties to allow the inflaton interactions with the Standard
Model particles for the reheating, since the interactions should be
sufficiently weak not to spoil the slow-roll inflation scenario.

Although the required suppression in Eq.~(\ref{suppression})
would be easily achieved by the finetuning, hereafter we will briefly 
discuss two possible resolutions of the finetuning problems.

\subsubsection{Multiple point principle ``explanation'' of suppression of
inflaton decay to Weinberg-Salam Higgses}
One possible attempt to explain the absence of the 
terms, $\eg$, $\lambda\,\abs{\phi_{\rm\sss WS}}^2 \, \phi^2$ 
term 
consists in calling upon a principle which has actually been 
developed in connection with the model discussed here as the 
example. This is the postulate of the multiple point 
principle (MPP)~\cite{MPP},
saying that the effective potential as a function of the 
scalars such as $\phi_{\sss\rm WS}$ and $\phi$ $\etc.$ should have 
many (as many as possible) equally deep minima. This is a 
principle that is actually true in superdsymmetric models in as far as 
there all the superdsymmetric states have to have zero 
and (thus) minimal energy density.

In the article~\cite{Froggatt:2001pa} it was argued that such a 
principle of equally deep minima would have a 
tendency to split into separate sectors -- totally 
decoupling ones -- once there are only one (or a few) 
free parameters
adjustable interactions between potentially separable sectors. 

Ignoring the irrelevant terms such as Eq.~(\ref{Lint})
the only interaction between an ``inflation sector'' 
and ``our sector'' -- meaning the Standard Model fields plus 
the seesaw neutrinos as well as possible gauge 
fields or scalars associated with it -- is supposed to be the 
term $\lambda\,\abs{\phi_{\rm\sss WS}}^2\,\abs{\phi}^2$ in 
Eq.~(\ref{Higgscouplings}).

Now we want to argue that provided we could find solutions for 
the coupling constants with, say, two degenerate minima in both of the 
mentioned sectors, when they were separate, we can argue for that 
$\lambda=0$ is at least one possible solution for satisfying 
the MPP requirement. Let us argue for that by counting relations 
($\approx$ finetunings) between the coupling constants needed to 
achieve degenerate minima in effective potentials: to achieve 
$n$ degenerate minima in an effective potential $V_{\rm eff}$ -- it be 
a function of one or more scalar fields -- one needs to finetune 
$(n-1)$ couplings or parameters. Let us for example imagine that if we
had our sector alone MPP could tune in $2-1=1$ relation between the 
couplings and thereby achieve $2$ degenerate minima (in the Weinberg-Salam 
Higgs effective potential) and that one also by tuning $2-1=1$ 
coupling/parameter can achieve two generate minima in the inflaton 
effective potential, if that were alone. Then we can see {\em easily}
that by combining the two sets of coupling constants and taking 
further $\lambda=0$, we achieve $4=2\cdot2$ degenerate 
minima for the effective potential of the full model. This 
effective potential is (at least) a function of both $\phi$ 
and $\phi_{\rm\sss WS}$ and it has its degenerate minima 
in all the $(\phi,\, \phi_{\rm\sss WS})$-pairs which are 
obtained by combining the $\phi$-values from the degenerate 
minima in the $\phi$-potential alone and the 
$\phi_{\rm\sss WS}$-values from the generate minima 
of ``our sector'' alone. There are $2\cdot2=4$ such combinations 
$(\phi,\, \phi_{\rm\sss WS})$ and thus this special solution with 
$\lambda=0$ provides $4$ degenerate minima. {\it A priori} 
we expect that $4$ degenerate minima requires $4-1=3$ finetuning 
relations between the couplings/parameters. The proposed 
solution with $\lambda=0$ uses just $3$ finetunings 
in as far as we used one finetuning relations for each of the 
two sectors plus the finetuning $\lambda=0$. The 
proposed solution has therefore not used more finetunings to 
be fixed by MPP than what is expected to be needed anyway. 
Had we instead imagined that we had 
got arranged say $3$ degenerate 
minima in the two sectors separately we would have been able to 
produce a solution with $\lambda=0$ and $9$ generate 
minima, that would have needed only $2+2+1=5$ finetunings 
against the {\it a priori} expected $9-1=8$ in this case. So in this
case we would quite clearly get $\lambda=0$.
We see here the possibility for the MPP to produce decoupling (totally)
of only loosely coupled sectors, such as the inflaton one and ours.

The region in field space in which this kind of argument fixes 
$\lambda$ conceived of as renormalization group running 
with the fields $(\phi,\, \phi_{\rm\sss WS})$ is of course where 
$\abs{\phi}^2\,\abs{\phi_{\rm\sss WS}}^2$ is large. So 
{\it a priori} one may wonder if the running of $\lambda$
could let it be non-zero for some other field values. It is, 
however, rather easily seen that the $\beta$-function for 
$\lambda$ only obtain terms proportional to 
$\lambda$ itself. Thus once zero the coupling 
$\lambda$ will remain zero under the running.

\subsubsection{Supersymmetric Extension} 

One of the most natural ways to explain the huge mass hierarchy 
between the fundamental Planck scale and the electroweak scale
is to introduce an additional symmetry, supersymmetry (SUSY).
It can stabilize the electroweak scale against the dangerous
radiative corrections. Here we briefly discuss the SUSY extension
of the family gauge replicated model and also its non-thermal 
leptogenesis.

In the family gauge replicated model with SUSY, we have 
to introduce two Weinberg-Salam Higgs (chiral super) 
fields, say $\phi_{u\rm\sss WS}$ and $\phi_{d\rm\sss WS}$, 
in order give masses to both the up-type quarks and 
the down-type quarks, respectively. Further, to ensure the cancellation 
of gauge and mixed anomalies, we introduce mirror (chiral 
super) fields of our Higgses, say, $W'$, $T'$, $\rho'$, $\omega'$, and 
$\phi_{\rm\sss SS}'$, for which the gauge charges are as 
if they were complex conjugates of $W$, $T$, $\rho$, $\omega$ 
and $\phi_{\rm \sss SS}$. Then, in the SUSY models, 
the number of the (chiral super) fields for Higgs becomes doubled.

The first question we would like to ask is: ``Can we 
reproduce the successful fitting of the masses 
and mixing for fermions which we performed 
in section 3?''. The answer is yes, to the first approximation. From 
the gauge invariance ($\ie$, six $U(1)$'s) we can write down 
the fermion mass matrices as Eqs.~(\ref{M_U}), (\ref{M_D}), (\ref{M_E}) 
and (\ref{Mdirac}) by replacing 
$\sVEV{\phi_{\rm\sss WS}} \to\sVEV{\phi_{d\rm\sss WS}}$ and 
$\sVEV{\phi_{\rm\sss WS}^\dagger} \to\sVEV{\phi_{u\rm\sss WS}}$, 
and also by replacing the dagger Higgs fields to the 
mirror (chiral super) fields, $\eg$, $W^\dagger\to W'$, 
as we must do since the superpotential shall be holomorphic.
If we take $\sVEV{\phi_{u\rm\sss WS}} =\sVEV{\phi_{d\rm\sss WS}} $, 
the overall scale of the Dirac matrices is the same for 
the up-quarks, down-quarks, charged-leptons, and Dirac neutrinos,
but is suppressed by $1/\sqrt{2}$ comparing with the non-SUSY model
which is outside of our discussion. Furthermore, from the $D$-term flatness
conditions, we expect that the VEVs of our 
Higgs fields are the same as those of mirror fields, $\eg$, 
$\sVEV{W} = \sVEV{W'}$. It should thus be noted that in this use 
we have only the five VEVs $\sVEV{W}=\sVEV{W'}$, $\sVEV{T}=\sVEV{T'}$, 
$\sVEV{\rho}=\sVEV{\rho'}$, $\sVEV{\omega}=\sVEV{\omega'}$ and 
$\sVEV{\phi_{\rm\sss SS}} = \sVEV{\phi'_{\rm\sss SS}}$ for the fitting
even though we had introduced the doubled Higgs fields in the model
compared with the non-SUSY model.
Therefore, the structure of the Dirac and Majorana mass matrices
remains unchanged at the cutoff scale, taken as the Planck scale.
If we neglect the effects of the evolution by the renormalization
group equation, we obtain the similar results given in Table~$2$.
Therefore, to the first approximation, we could say that 
we keep the successful fitting of masses and mixings 
even with SUSY introduced into the model.

The non-thermal leptogenesis is found to work in the SUSY version.
By introducing SUSY, the total decay rate of the inflaton $\Gamma_\phi$
through the interaction Eq.~(\ref{Lint}),
the decay rate asymmetry parameters of Majorana neutrinos $\epsilon_i$,
and the degree of freedom of the thermal bath of the universe
change only within a factor two which is also out of our discussion.
Then, the inflaton of mass $m_\phi \sim 2 M_{2,3}$ induce 
the reheating temperature of $T_R \sim 10^{6} \GeV$  and hence
the correct order of baryon asymmetry is produced via the proposed scheme.
Interestingly, with  such low reheating temperatures we can avoid the 
cosmological gravitino problem~\cite{GravitinoP}.

Finally, we would like to ask; ``Can we suppress the interactions
between the Weinberg-Salam Higgs fields and the inflaton?''.
The potentially dangerous interactions would be induced from the term
in the superpotential as ${\cal W} = \phi \phi_{u\rm\sss WS} 
\phi_{d\rm\sss WS}$.
This term can be killed by using the chiral $U(1)_R$ symmetry
and the global $U(1)_X$ symmetry.
We assign the $R$-charges $0$ for the Higgs fields and the inflaton,
and $1$ for the matter fields.
On the other hand, we assign the $X$-charges $0$ for our Higgs fields 
($W$, $T$, $\rho$ and $\omega$) and 
the inflaton, $-2x$ for two Weinberg-Salam Higgs fields
and the seesaw Higgs, and $x$ 
for the quark and lepton fields ($x$ is the charge corresponding to $U(1)_X$).
This symmetry forbids ${\cal W} = \phi \phi_{u\rm\sss WS} \phi_{d\rm\sss WS}$
as well as the $\mu$-term ${\cal W} = \mu \phi_{u\rm\sss WS} 
\phi_{d\rm\sss WS}$. 
The required $\mu$-term may be generated by the effect of the 
SUSY breaking. Therefore, we can avoid the fast decay of 
the inflaton into the Weinberg-Higgs fields. Notice that 
the dangerous proton decay induced by the dimension 
five operator is also absent due to this $U(1)_R$ symmetry.

As we had briefly sketched, in the family replicated gauge group with
SUSY, we have, to the first approximation, the successful fitting of
masses and mixings for fermions as well as the non-thermal
leptogenesis via heavier Majorana neutrinos produced by the late
decays of the inflatons.

\subsection{Further generalization, abstracting the model} 

It should be remarked that the idea of how to get the 
heavier seesaw neutrinos produced non-thermally in 
decay of an inflation boson, could be generalized to 
letting the boson -- replacing the $\phi$ boson above -- 
be any sufficiently long lived boson at the end decaying 
dominantly into the (heavier) seesaw neutrinos. Then 
it is possible that at the stage of the 
cosmological development when the temperature 
passes the mass $m_\phi$ of these particles, 
they become non-relativistic and we obtain a 
matter dominated era. Such an era will end when 
Hubble expansion rate, $H$, becomes of the order 
of the width $\Gamma_{\phi}$ by the decay of 
these particles -- pretty analogous to 
the ``reheating'' talked about above.

That it is a true inflaton does not really matter, we 
could get by the same calculation as we did the 
same leptogenesis anyway. In this way we could escape 
completely mass estimates suggested for the 
inflaton. However, since the inflaton mass is very 
strongly dependent on the form of the inflation
effective potential used, there is really no general mass 
estimation for the inflaton (in a model independent 
way). Therefore, it was correct that we used the mass for fitting 
$m_\phi{}\sim{}2M_{2,3}{}\sim10^{10}~\GeV$.
 
\section{Conclusions} 
\label{sec6}

In this article the leptogenesis by the decays of 
the heavier Majorana neutrinos $N_2$ and $N_3$ was considered.
We pointed out that the $N_2$ and $N_3$ decays could be 
responsible for generating the baryon 
asymmetry in the present Universe,
if they are produced non-thermally in the inflaton decays.
In our picture then the reheating temperature of 
the inflation is sufficiently low that
the wash-out processes mediated by the lighter Majorana neutrino 
are ineffective.

Accepting this picture for baryogenesis would release the family 
replicated gauge group model from its only severe deviation from 
agreement with experiment -- the amount of baryon minus lepton number 
produced in the seesaw era -- and thus make it agree within its 
pretended only order of magnitude accuracy with all quantities 
it predicts.

This means that with the combined picture of an inflaton 
as suggested here and the family 
replicated gauge group model -- in the latest version -- we have
a very viable model where we used the five parameters 
to produce the fit of Table~$2$ and some adjustment of the 
inflation properties essentially all the parameters useful for 
going beyond the Standard Model, however, only order of 
magnitudewise. Indeed we have now fit all the quark and 
lepton masses and mixings -- including the neutrinos and 
the $CP$ violation -- and the baryon asymmetry, and 
further successfully coped with bounds on proton decay and lepton flavor 
violating decays. Also bounds on the neutrinoless beta 
decay are respected though so that neither the positive 
findings of the Heidelberg-Moscow collaboration 
of $\sVEV{\abs{m_{ee}}}\approx0.39~\eV$~\cite{klapdor} 
nor the LSND neutrino~\cite{LSND} are welcome in our scheme.

\section*{Acknowledgments}

We wish to thank Prof.~T.~Yanagida for useful discussions, 
especially, pointing out the possibility of having the low reheating 
temperature and the use of Eq.~(\ref{Lint}) discussed in Section $4$.
One of us (Y.T.) expresses his gratitude to Prof.~J.~Wess
for warm hospitality at Universit{\"a}t M{\"u}nchen extended 
to him during his visit where part of this work was done.
H.B.N. thanks the Alexander von Humboldt-Stiftung for the Forschungspreis. 
Y.T. thanks DESY for financial support, and the Frederikke 
L{\o}rup f{\o}dt Helms Mindelegat for a travel grant to 
attend the Neutrino 2002, Munich, Germany. 
%

\end{document}